\documentclass[cite]{epl}%
\usepackage{amsmath}
\usepackage{amssymb}
\usepackage{graphicx}%
\usepackage{amsfonts}
\shorttitle{Dynamical Spin-Blockade in a quantum dot ...}
\institute{
Departement Physik und Astronomie, Universit\"at Basel,
Klingelbergstrasse~82,
CH-4056 Basel, Switzerland}
\begin{document}

\title{Dynamical Spin-Blockade in a quantum dot with paramagnetic leads}
\author{A. Cottet and W. Belzig}
\maketitle

\begin{abstract}
We investigate current fluctuations in a three-terminal quantum dot in the
sequential tunneling regime. Dynamical spin blockade can be induced when the
spin-degeneracy of the dot states is lifted by a magnetic field. This results
in super-Poissonian shot noise and positive zero-frequency cross-correlations.
Our proposed setup can be realized with semiconductor quantum dots.

\end{abstract}


\vspace*{-5mm}
\section{Introduction}

Non-equilibrium current noise in mesoscopic structures is a consequence of the
discreteness of the charge carriers (for reviews, see
Refs.~\cite{blanter:00,nazarov:03}). For conductors with open channels the
fermionic statistics of electrons results in a suppression of shot noise below
the classical Schottky limit \cite{schottky}. This was first noted by Khlus
\cite{khlus:87} and Lesovik \cite{lesovik:89} for single channel conductors.
Subsequently, B\"{u}ttiker generalized this suppression for many-channel
conductors \cite{buettiker:90}. Mesoscopic conductors are often probed by two
or more leads. The quantum statistics induces cross-correlations between the
currents in different terminals. Since these cross-correlations vanish in the
classical limit, even their sign is not obvious a priori. Using only the
unitarity of the scattering matrix, B\"{u}ttiker proved that
cross-correlations for non-interacting fermions are \textit{always negative}
for circuits with leads maintained at constant potentials \cite{buettiker:92}.
Note that this also holds in the presence of a magnetic field. It has also
been found that an interacting paramagnetic dot shows negative
cross-correlations in the absence of a magnetic field \cite{bagrets:02}.
Spin-dependent cross-correlations in a non-interacting 4-terminal spin valve
were studied \cite{belzig:03} and found to be negative. On the experimental
side negative cross-correlations were measured by Henny \textit{et al.}
\cite{henny:99,oberholzer:00} and Oliver \textit{et al.} \cite{oliver:99} in
mesoscopic beam splitters.

Several ways to produce positive cross-correlations in fermionic systems have
been proposed (see e.g. \cite{buettiker:03-book} for a recent review). Among
these possibilities are sources which inject correlated electrons
\cite{p1,p2,p3,p4,p5,schechter,p6,p7,p8,p9,p10,p11,p12,p13,p14} and finite-frequency
voltage noise \cite{martin:00,buettiker:03-book}. The question of the
existence of intrinsic mechanisms, i.~e. due to interactions occuring in the
beam-splitter device itself, has been answered positively by us
\cite{audrey:03}. Surprisingly, a simple quantum dot connected to
ferromagnetic contacts can lead to positive cross-correlations due the
so-called \textit{dynamical spin-blockade}. Simply speaking, up- and
down-spins tunnel through the dot with different rates. In the limit where the
Coulomb interaction prevents a double occupancy of the dot, the spins which
tunnel with a lower rate modulate the tunneling of the other spin-direction,
which leads to an effective \textit{bunching} of tunneling events. In a three
terminal geometry with one input and two outputs, this results in positive
cross-correlation between the two output currents. Independently, Sauret and
Feinberg proposed a slightly different setup of a ferromagnetic quantum dot,
which also produces positive cross-correlations \cite{sauret:03}.

Experimentally, it is more difficult to fabricate quantum dots with
ferromagnetic leads. However, quantum dots with paramagnetic leads have shown
to exhibit spin-dependent transport. A magnetic field lifts the
spin-degeneracy and a spin-polarized current with nearly 100\% efficiency can
be created \cite{hanson:03a,Recher}. In this Letter, we will address the
current correlations in a few-electron quantum dot connected to three
paramagnetic leads. We will show below that positive cross-correlations can be
produced in this device simply by applying a \textit{magnetic field}.
Furthermore, this system also shows a super-Poissonian shot noise.

To arrive at these conclusions we consider a quantum dot with one orbital
energy level $E_{0}$ connected to three terminals by tunnel contacts. The
junctions are characterized by bare tunneling rates $\gamma_{i}$ ($i=1,2,3$)
and capacitances $C_{i}$. We assume that a magnetic field $B$ is applied to
the dot, which leads to a Zeeman splitting of the level according to
$E_{\downarrow(\uparrow)}=E_{0}+(-)g\mu_{B}B/2$, where $\mu_{B}=e\hbar/2m $ is
the Bohr magneton. The double occupancy of the dot costs the charging energy
$E_{c}=e^{2}/2C$, with $C=\sum_{i}C_{i}$. The energy spacing to the next
orbital is $\Delta$. We will assume
\begin{equation}
k_{B}T,eV,\mu_{B}B\ll E_{c},\Delta\,.\label{eq:parameters}%
\end{equation}
According to these inequalities, the dot can be only singly occupied and we
have to take into account only one orbital level.


In the sequential-tunneling limit $\hbar\gamma_{j}\ll k_{B}T$, the time
evolution of the occupation probabilities $p_{\psi}(t)$ of states $\psi
\in\{\uparrow,\downarrow,0\}$ is described by the master equation:
\begin{equation}
\frac{d}{dt}p_{\psi}=M_{\psi\varphi}p_{\varphi}\label{MasterEquation}%
\end{equation}
where
\begin{equation}
\hat{M}=\left[
\begin{array}
[c]{ccc}%
-\Gamma_{\uparrow}^{-}-\Gamma_{\downarrow\uparrow} & \Gamma_{\uparrow
\downarrow} & \Gamma_{\uparrow}^{+}\\
\Gamma_{\downarrow\uparrow} & -\Gamma_{\downarrow}^{-}-\Gamma_{\uparrow
\downarrow} & \Gamma_{\downarrow}^{+}\\
\Gamma_{\uparrow}^{-} & \Gamma_{\downarrow}^{-} & -\Gamma_{\uparrow}%
^{+}-\Gamma_{\downarrow}^{+}%
\end{array}
\right]  \,.\label{MatrixM}%
\end{equation}
The rate for an electron to tunnel on/off the dot ($\epsilon=+/-$) through
junction $j$ is given by $\Gamma_{j\sigma}^{\epsilon}=\gamma_{j}%
/(1+\exp[\epsilon(E_{\sigma}-eV_{j})/k_{B}T])$, where $V_{1}=V_{3}=-C_{2}V/C$
and $V_{2}=(C_{1}+C_{3})V/C$. Here, we took the Fermi energy $E_{F}=0$ for
lead 2 as a reference. The total tunneling rates are $\Gamma_{\sigma
}^{\epsilon}=\sum_{j}\Gamma_{j\sigma}^{\epsilon}$ and $\gamma=\sum_{j}%
\gamma_{j}$. Spin flips on the dot are described by rates $\Gamma
_{\downarrow\uparrow(\uparrow\downarrow)}$, which obey the detailed balance
rule $\Gamma_{\uparrow\downarrow}/\Gamma_{\downarrow\uparrow}=\exp(g\mu
_{B}B/k_{B}T)$. From Eq.~(\ref{MasterEquation}) the stationary occupation
probabilities $\bar{p}_{\sigma}$ are
\begin{equation}
\bar{p}_{\sigma}=\frac{\Gamma_{\sigma}^{+}\Gamma_{-\sigma}^{-}+\Gamma
_{\sigma,-\sigma}(\Gamma_{\uparrow}^{+}+\Gamma_{\downarrow}^{+})}{\gamma
^{2}-\Gamma_{\uparrow}^{+}\Gamma_{\downarrow}^{+}+(\gamma+\Gamma_{\downarrow
}^{+})\Gamma_{\uparrow\downarrow}+(\gamma+\Gamma_{\uparrow}^{+})\Gamma
_{\downarrow\uparrow}}\;,
\end{equation}
and $\bar{p}_{0}=1-\bar{p}_{\uparrow}-\bar{p}_{\downarrow}$. These
probabilities can be used to calculate the average value $\langle I_{j}%
\rangle$ of the tunneling current $I_{j}(t)$ through junction $j$ as
\begin{equation}
\langle I_{j}\rangle=e\sum_{\epsilon,\sigma}\epsilon\Gamma_{j\sigma}%
^{\epsilon}\bar{p}_{A(\sigma,-\epsilon)}\,,\label{eq:current}%
\end{equation}
where $A(\sigma,\epsilon)$ is the state of the dot after the tunneling of an
electron with spin $\sigma$ in the direction $\epsilon$, i.~e., $A(\sigma
,-1)=0$ and $A(\sigma,+1)=\sigma$. The frequency\ spectrum of the noise
correlations can be defined as
\begin{equation}
S_{ij}(\omega)=2\int_{-\infty}^{+\infty}dt\exp(i\omega t)\langle\Delta
I_{i}(t)\Delta I_{j}(0)\rangle\,,\label{CorrelationsDefinition}%
\end{equation}
where $\Delta I_{i}(t)=I_{i}(t)-\langle I_{i}\rangle$ is the deviation from
the average current in terminal $i$. It can be calculated using the method
developed in Refs. \cite{setnoise1,setnoise2,setnoise3} as:%
\begin{equation}
\frac{S_{ij}(\omega)}{2e^{2}}=\sum_{\epsilon,\sigma}\Gamma_{j\sigma}%
^{\epsilon}\bar{p}_{A(-\epsilon,\sigma)}\delta_{ij}+\sum_{\sigma\sigma
^{\prime}\epsilon\epsilon^{\prime}}\epsilon\epsilon^{\prime}S_{i\sigma
j\sigma^{\prime}}^{\epsilon\epsilon^{\prime}}(\omega)\;,\label{eq:noise1}%
\end{equation}
where the first term is the Schottky noise produced by tunneling through
junction $j$, and
\begin{equation}
S_{i\sigma j\sigma^{\prime}}^{\epsilon\epsilon^{\prime}}(\omega)=\Gamma
_{i\sigma}^{\epsilon^{\prime}}G_{A(\sigma,-\epsilon^{\prime}),A(\sigma
^{\prime},\epsilon)}(\omega)\Gamma_{j\sigma^{\prime}}^{\epsilon}\bar
{p}_{A(\sigma^{\prime},-\epsilon)}+\Gamma_{{j}\sigma^{\prime}}^{\epsilon
^{\prime}}G_{A(\sigma^{\prime},-\epsilon^{\prime}),A(\sigma,\epsilon)}%
(-\omega)\Gamma_{i\sigma}^{\epsilon}\bar{p}_{A(\sigma,-\epsilon)}%
\,,\label{eq:noise2}%
\end{equation}
with $G_{\psi,\varphi}(\omega)=\bar{p}_{\psi}/i\omega-\left(  i\omega
+M\right)  _{\psi,\varphi}^{-1}$.


In the following we study the dot in a beam-splitter configuration, in which a
bias voltage $V$ is applied between terminal 2 and terminals 1 and 3. We
consider the case $V>0$, so that it is energetically more favorable for
electrons to go from lead 2 to leads 1 and 3. We will limit our discussion to
the case in which the two Zeeman sublevels are below the Fermi energy at
equilibrium (\textit{i.~e.} $E_{0}\pm g\mu_{B}B/2<0$). The opposite case was
discussed in Ref.~\cite{thielmann:03} for a two-terminal dot. We are mostly
interested in the total zero-frequency current noise $S_{22}=S_{22}(0)$ and
the cross-correlations $S_{13}=S_{13}(0)$ between the two output leads. It is
useful to define the Fano factor $F_{2}=S_{22}/2e\langle I_{2}\rangle$ and,
correspondingly, $F_{13}=S_{13}/2e \langle I_{2}\rangle$.

\begin{figure}[t]
\centering\includegraphics[width=10cm,keepaspectratio,clip]{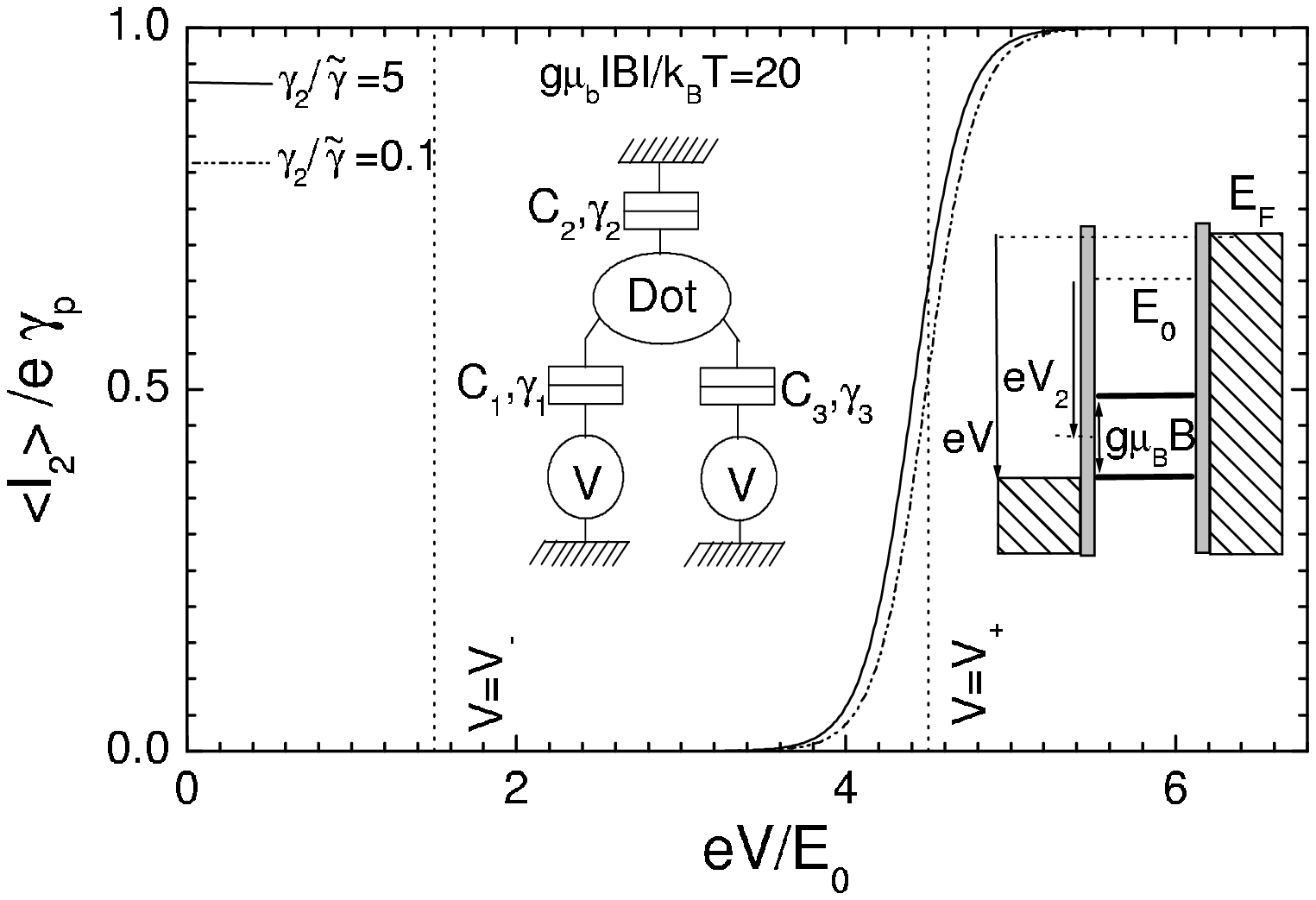}
\caption{Current-voltage characteristic of the circuit shown in the inset for
$E_{0}<0$, $C_{1}=C_{2}=C_{3}$, $\gamma_{1}=\gamma_{3}$, $k_{B}T/\left\vert
E_{0}\right\vert =0.05$, $g \mu_{b}B/\left\vert E_{0}\right\vert =1$, and
different values of $\gamma_{2}/\widetilde{\gamma}$. The average current
$\langle I_{2}\rangle$ through lead $2$ is plotted in units of $e \gamma
_{p}=2e\gamma_{2}\tilde\gamma/(\tilde\gamma+2\gamma_{2})$; the voltage is in
units of $E_{0}$. The positions of $V_{+}$ and $V_{-}$ are indicated in dotted
lines. }%
\label{fig:current}%
\end{figure}

In the following, we will first assume that $k_{B}T\ll g\mu_{B}B$. Transport
through the down level is energetically allowed for $V\gtrsim V_{-}%
=(-E_{0}-g\mu_{B}B/2)C/eC_{2}$. However, for $V\lesssim V_{+}=(-E_{0}+g\mu
_{B}B/2)C/eC_{2}$, the dot is blocked by an up spin, thus down spins cannot
cross the dot. Around $V\simeq V_{+}$, the lower Zeeman level is close to the
Fermi level of leads 1 and 3, as represented by the level diagram in the lower
right inset of Fig.~\ref{fig:current}. The blockade of the dot is then
partially lifted and transport through both levels is allowed. In this regime,
we can write the tunneling rates as $\Gamma_{2\sigma}^{+}=\gamma_{2}$,
$\Gamma_{2\sigma}^{-}=0$, $\Gamma_{1(3)\uparrow}^{-}=x\gamma_{1(3)}$,
$\Gamma_{1(3)\uparrow}^{+}=(1-x)\gamma_{1(3)}$, $\Gamma_{1(3)\downarrow}%
^{+}=0$, and $\Gamma_{1(3)\downarrow}^{-}=\gamma_{1(3)}$, where $x=1/(1+\exp
[-(E_{0}-g\mu_{B}B-eV_{1})/k_{B}T])$ ranges from 0 to 1 with increasing
voltage. Furthermore, taking $\gamma_{sf}=0$ and $\tilde{\gamma}=\gamma
_{1}+\gamma_{3}$, we find for the current
\begin{equation}
\langle I_{2}\rangle=\frac{2ex\gamma_{2}\tilde{\gamma}}{\tilde{\gamma}%
+\gamma_{2}(1+x)}\,,\label{I2para}%
\end{equation}
for the Fano factor
\begin{equation}
F_{2}=1+\frac{2\gamma_{2}\left(  \tilde{\gamma}(1-3x)+(1-x)^{2}\gamma
_{2}\right)  }{\left(  \tilde{\gamma}+\gamma_{2}(1+x)\right)  ^{2}%
}\,,\label{F2para}%
\end{equation}
and for the cross correlations
\begin{equation}
F_{13}=\frac{\gamma_{1}\gamma_{3}}{\tilde{\gamma}^{2}}\frac{2(1-x)^{2}%
\gamma_{2}^{3}+(1-7x+x^{2}+x^{3})\gamma_{2}^{2}\tilde{\gamma}-2(1-x^{2}%
)\gamma_{2}\tilde{\gamma}^{2}-(1-x)\tilde{\gamma}^{3}}{\gamma_{2}%
(\tilde{\gamma}+(1+x)\gamma_{2})^{2}}\,.\label{eq:C2para}%
\end{equation}
We observe that the current increases with voltage (i.e. with $x$) around the
voltage step $V_{+}$. Note that this current is \textit{not spin-polarized}
because up and down spin have the same probability to enter the dot,
regardless of what happens at the output. The Fano factor $F_{2}$ and the
cross-correlations $F_{13}$ deviate from their Poissonian values depending on
the applied voltage. Our main results are that $F_{2}$ can be super-Poissonian
and $F_{13}$ positive for $x<1$, as can be cleary seen from (\ref{F2para}) or
(\ref{eq:C2para}) in the limit $\gamma_{2}\gg\tilde{\gamma}$. These features
are a consequence of dynamical spin blockade: up spins leave the dot with a
rate smaller than down spins, leading to a bunching of tunneling events
\cite{audrey:03}. In the limit $x\rightarrow1$, the Fano factor is always
sub-Poissonian and the cross-correlations always negative. This is due to the
fact\ that the tunnel rates of up and down spins are equal, thus the Zeeman
splitting plays no role and the dot is equivalent to a simple quantum dot with
a spin-degenerate level. In the limit $x\rightarrow0$, one could also expect
the super-Poissonian nature of $F_{2}$ and the positivity of $F_{13}$ to be
lost since the transport is enabled only by thermally activated processes.
However, below the voltage threshold $V_{+}$, the Fano factor tends to:
\begin{equation}
F_{2}=1+\frac{2\gamma_{2}}{\gamma_{2}+\tilde{\gamma}}\,.
\end{equation}
which is always super-Poissonian. If the coupling to terminal 2 dominates,
i.~e. $\gamma_{2}\gg\tilde{\gamma}$, the Fano factor takes a maximal value of
3. In the opposite limit $\tilde{\gamma}\gg\gamma_{2}$, $F_{2}$ approaches the
Poisson limit of uncorrelated single charge transfer. It is interesting to
note that a symmetric junction $\tilde{\gamma}=\gamma_{2}$ produces twice the
Poisson noise level. The cross-correlations in the same limit have the form
\begin{equation}
F_{13}=\frac{\gamma_{1}\gamma_{3}}{\tilde{\gamma}^{2}}\frac{(2\gamma
_{2}+\tilde{\gamma})(\gamma_{2}-\tilde{\gamma})}{\gamma_{2}(\gamma_{2}%
+\tilde{\gamma})}\,.\label{eq:cross_below_step}%
\end{equation}
In the three cases discussed above, the cross-correlations thus take the
limiting values $F_{13}=-\gamma_{1}\gamma_{3}/\gamma_{2}\tilde{\gamma}$ for
$\tilde{\gamma}\gg\gamma_{2}$, $F_{13}=0$ for $\tilde{\gamma}=\gamma_{2}$ and
$F_{13}=2\gamma_{1}\gamma_{3}/\tilde{\gamma}^{2}$ for $\gamma_{2}\gg
\tilde{\gamma}$. Hence, both the super-Poissonian nature of $F_{2}$ and the
positivity of $F_{13}$ can persist for $x\rightarrow0$. Even if the transport
is enabled only by thermally activated processes, dynamical spin blockade
already results in a correlated transfer of electrons.

\begin{figure}[ptbh]
\centering\includegraphics[width=\textwidth,keepaspectratio,clip]{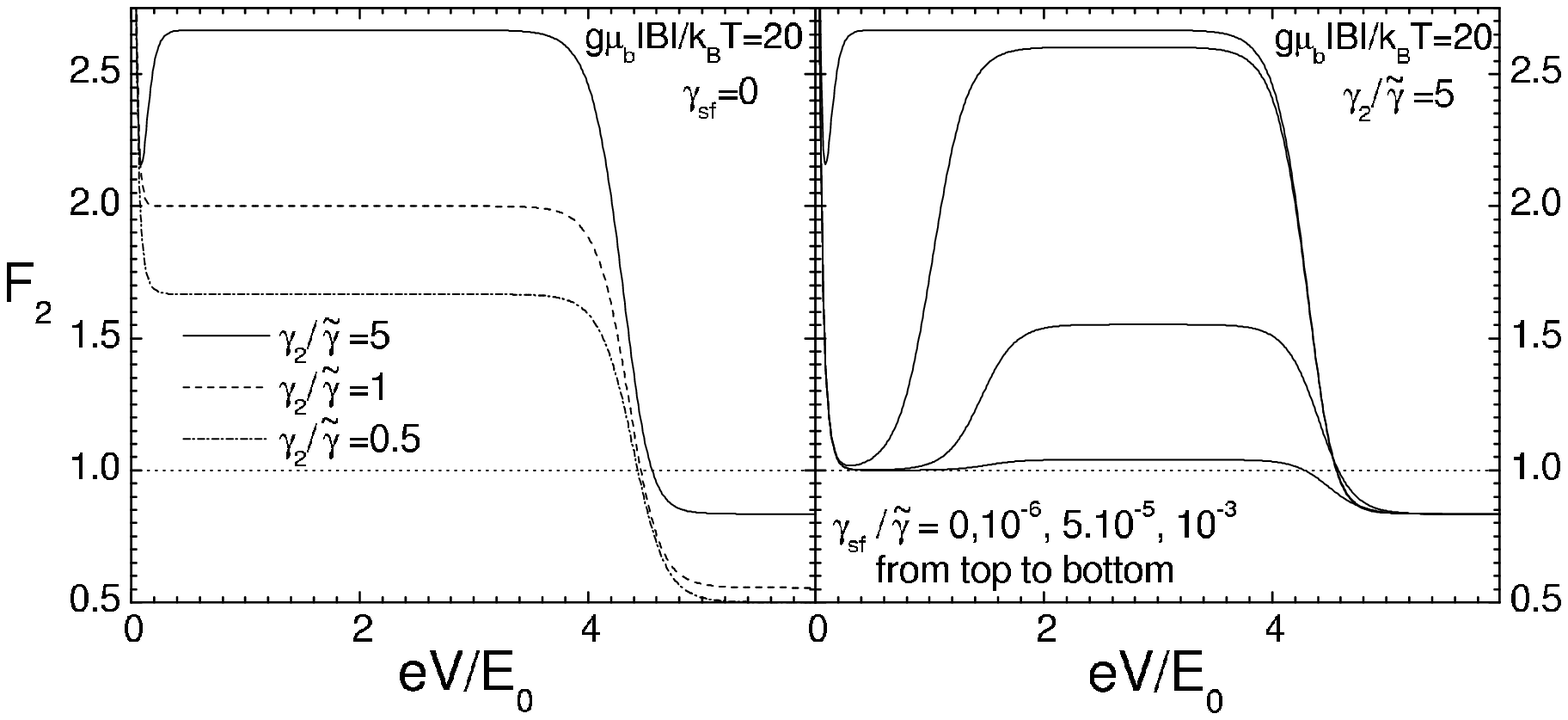}\caption{Fano
factor $F_{2}=S_{22}/2e\langle I_{2}\rangle$ of the total current as a
function of voltage, for the same circuit parameters as in
Fig.~\ref{fig:current}. Left panel: Data for different values of $\gamma
_{2}/\tilde{\gamma}$ and $\gamma_{sf}=0$. Right panel: Effect of spin flip
scattering for $\gamma_{2}/\widetilde{\gamma}=5$ and different values of
$\gamma_{sf}$. The curves displayed in both panels are independent of the
asymmetry between the output leads.}%
\label{fig:fano}%
\end{figure}

\begin{figure}[ptbh]
\centering\includegraphics[width=72mm,keepaspectratio,clip]{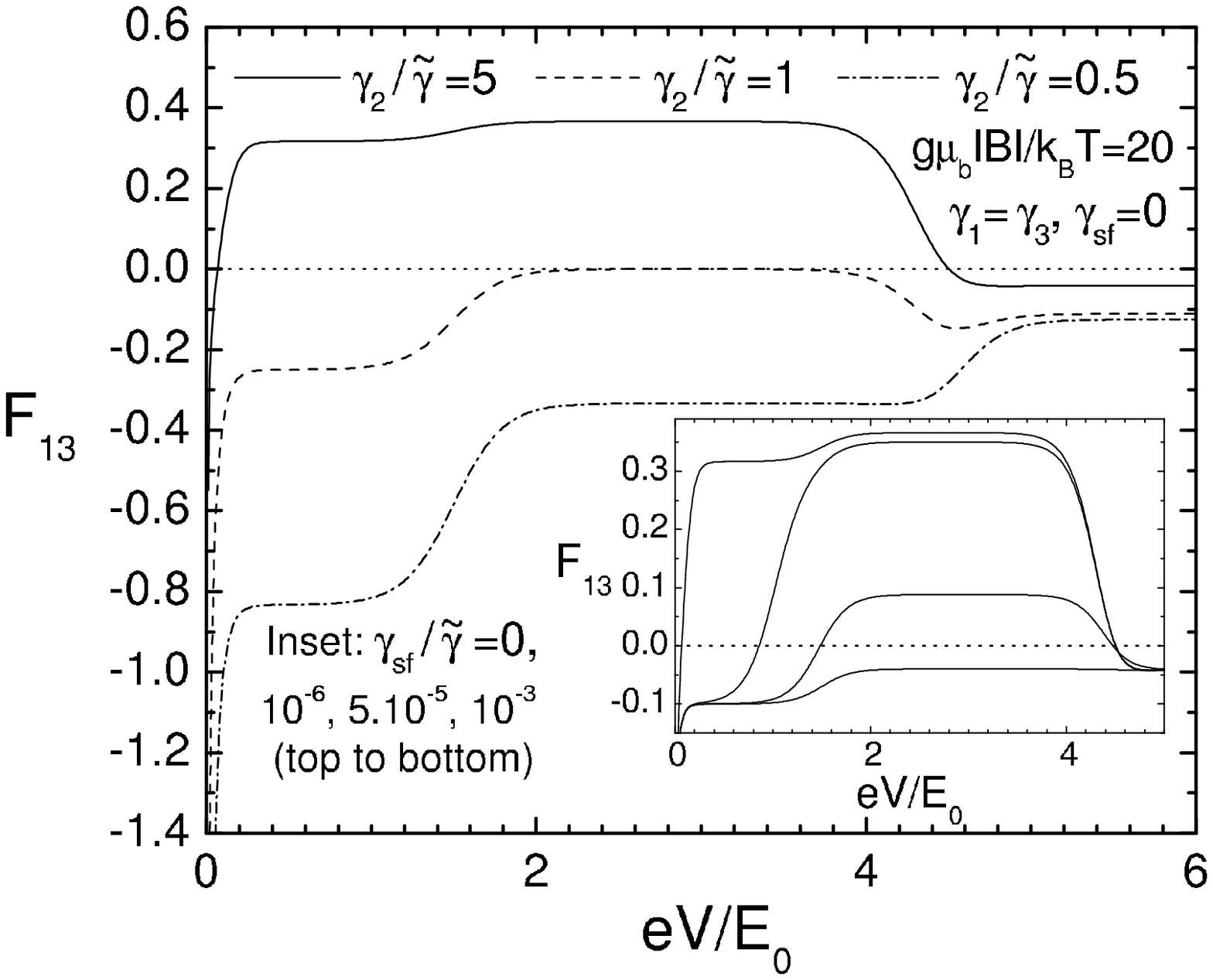}
\hspace{1mm}%
\includegraphics[width=68mm,keepaspectratio,clip]{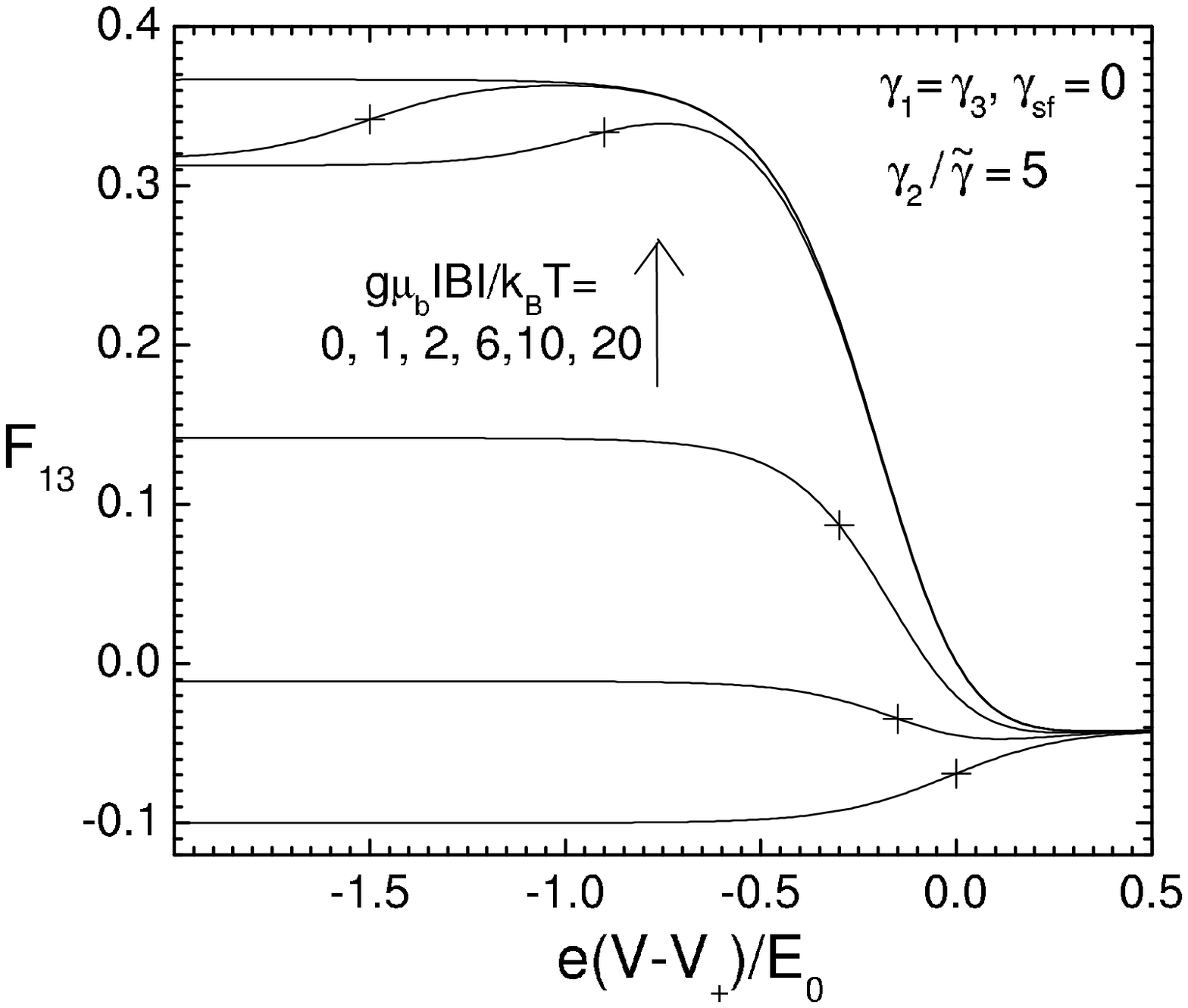}\caption{Left:
Cross-Fano factor $F_{13}=S_{13}/2e\langle I_{2}\rangle$ between leads 1 and 3
as a function of voltage, for the same circuit parameters as in Fig.
\ref{fig:current} and different values of $\gamma_{2}/\widetilde{\gamma}$. In
all curves of the main frame $\gamma_{sf}=0.$ The inset shows the effect of
spin flip scattering for $\gamma_{2}/\widetilde{\gamma}=5$ and different
values of $\gamma_{sf}$. Right: $F_{13}$ as a function of voltage for
different values of $B$. The curves are shown for $\gamma_{2}/\widetilde
{\gamma}=5$. A cross indicates the position of $V_{-}$ for each case.}%
\label{fig:cross}%
\end{figure}

We now turn to the discussion of the general results displayed in
Figs.~(\ref{fig:current})-(\ref{fig:cross}), obtained from an exact treatment
of the full Master equation. Fig.~\ref{fig:current} shows the full voltage
dependence of the average current. As expected, the current shows a single
step at $V\approx V_{+}$
\cite{ralph:95,ralph:97,cobden:98,cobden:02,hanson:03}. The step width is
about $10k_{B}TC/eC_{2}$, whereas its position varies only slightly with the
asymmetry of the junctions (the maximal variation is about $0.7k_{B}TC/eC_{2}$).

The left panel of Figure \ref{fig:fano} shows the voltage dependence of the
Fano factor in the absence of spin-flip scattering, for some values of
$\gamma_{2}/\tilde{\gamma}$. The divergence $2k_{B}T/eV$ of the Fano factor at
zero voltage is simply a result of the dominating thermal noise in the limit
$k_{B}T>eV$. Note that similarly to $\langle I_{2}\rangle$, the Fano factor
$F_{2}$ shows one single step at $V\sim V_{+}$. The right panel of
Fig.~\ref{fig:fano} shows the effect of spin-flip scattering on $F_{2},$ for
the case $\gamma_{2}=5\tilde{\gamma}$. For $V_{-}<V<V_{+}$ spin flips become
effective when $\Gamma_{\uparrow\downarrow}=\gamma_{sf}\exp(g\mu_{B}%
B/2k_{B}T)\sim\gamma_{i}$, see Eq.~(\ref{MatrixM}). The sensitivity to
$\gamma_{sf}$ thus increases with B. Below $V_{-}$, even smaller spin-flip
rates suppress the super-Poissonian noise because the dwell time of electrons
on the dot is very long. Far above $V_{+}$, spin-flip scattering has no effect
on the sub-Poissonian noise.

The left panel of Fig.~\ref{fig:cross} shows the voltage dependence of the
cross-correlations factor $F_{13}$ between the two output terminals, for the
same parameters as in Fig.~\ref{fig:fano}. First, around the voltage threshold
$V_{+}$, we observe the features discussed above. The cross-correlations
develop from a positive or negative level below $V_{+}$ depending on the ratio
$\gamma_{2}/\tilde{\gamma}$ to the usual negative cross-correlations above
$V_{+}$, where the spin-splitting plays no role anymore. Remarkably, in
contrast to $F_{2}$, the cross-correlations also show a step around the lower
voltage threshold $V_{-}$. This illustrates clearly that $F_{13}$ and $F_{2}$
are qualitatively different. The absence of the lower step for $F_{2}$ can be
interpreted as a consequence of the unidirectionnality of tunneling through
junction 2. Indeed, $\Gamma_{2\sigma}^{-}\rightarrow0$ means that $F_{2}$
depends only on $\bar{p}_{0} $ and $G_{0,\uparrow(\downarrow)}$ [see
(\ref{eq:current}) and (\ref{eq:noise2})]. Now, for $V\sim V_{-},$
$\Gamma_{1/3\uparrow}^{-}\rightarrow0$ implies that the contribution of these
terms is independent of $V$. On the contrary, $F_{13}$ also depends on
$\bar{p}_{\uparrow(\downarrow)}$ and $G_{\sigma,0}$ with $\sigma\in
\{\uparrow,\downarrow,0\}$. For $\Gamma_{1/3\uparrow}^{-}\rightarrow0$, these
last terms depend strongly on $\Gamma_{1/3\downarrow}^{-}$ which varies itself
significantly with $V$ around $V_{-}$. Note that the absence of a step in
$F_{2}$ implies a redistribution of the noise between $S_{11}$, $S_{33}$ and
$S_{13}$ when the threshold $V_{-}$ is crossed (due to charge conservation,
$S_{22}=S_{11}+S_{33}+2S_{13}$). The extra step of $F_{13}$ disappears for
$\gamma_{2}\gg\tilde{\gamma}$. In this limit, the cross-correlations display a
single low voltage plateau $F_{13}=2\gamma_{1}\gamma_{3}/\tilde{\gamma}^{2}$,
which is an upper bound for the two low voltages plateaux found in the general
case. The inset in the left panel of Fig.~\ref{fig:cross} shows the effect of
spin-flip scattering on the cross-correlations. As expected, they suppress all
spin-effects and the positive cross-correlations become finally negative. Like
for the Fano factor, very small spin-flip scattering rates $\gamma_{sf}$ are
already sufficient to modify $F_{13}$ for $V<V_{+}$.

Since the positive cross-correlations found in this work are intimately
related to the dynamical spin-blockade, we expect a strong dependence on the
magnetic field. The right panel of Fig.~\ref{fig:cross} shows the voltage
dependence of $F_{13}$ around the step $V_{+}$, for a fixed temperature and
various magnetic fields. Just below $V_{+}$ the limiting value of $F_{13}$ is
determined by formula (\ref{eq:cross_below_step}). Thus, for a constant
voltage $V\leq V_{+}$ we predict a cross-over from negative to positive
cross-correlations with increasing magnetic field. One can see a qualitative
change in the curves, which can be understood by a gradual splitting of the
voltage steps $V_{-}$ and $V_{+}$. The lower step is at $V_{-}-V_{+}=-g\mu
_{B}BC/C_{2}e$. As long as $g\mu_{B}B\lesssim10k_{B}T$, the two voltage steps
are indistinguishable. However, positive cross-correlations are already
expected for $g\mu_{B}B\gtrsim2k_{B}T$. For $g\mu_{B}B=6k_{B}T$ the two steps
still overlap, resulting in a broad peak, whereas for the higher magnetic
field $g\mu_{B}B=20k_{B}T$ the lower threshold at $V_{-}$ is outside the
plotting region.

The regime $V\sim V_{+}$ has the advantage that the current is not
exponentially small (c.f. Fig.~\ref{fig:current}) and thus observable more
easily in an experiment. For $\gamma_{1}=\gamma_{2}/5=\gamma_{3}$, the maximum
value obtained for the cross-correlations is $S_{13}\simeq0.09e^{2}\gamma_{p}$
at $x\simeq0.17$, i.~e. $V\simeq4.26 E_{0}$ in Fig.~\ref{fig:cross}. With
$\gamma_{p}\simeq5$ GHz, this corresponds to $10^{-29}$A$^{2}$s, a noise level
accessible experimentally \cite{birk:95}.

In conclusion we have studied current correlations for a three terminal
quantum dot with unpolarized leads, placed in a magnetic field. Below the
voltage threshold $V_{+}$, as a result of dynamical spin-blockade, the Fano
factor of the input current shows an interesting super-Poissonian behavior and
the cross-correlations in the two output leads can be positive. At higher
voltages the Fano factor becomes sub-Poissonian and the cross-correlations
negative, as usual. The effect we predict should be observable in
semiconductor quantum dots of Ref.~\cite{hanson:03}.

\acknowledgments

We acknowledge discussions with C. Bruder, H.-A. Engel and T. Kontos. This
work was financially supported by the RTN Spintronics, the Swiss NSF and the
NCCR Nanocience.

\end{document}